# Voltage modulated electro-luminescence spectroscopy and negative capacitance – the role of sub-bandgap states in light emitting devices

Kanika Bansal and Shouvik Datta

Division of Physics, Indian Institute of Science Education and Research, Pune 411021, Maharashtra, India

## Abstract

Voltage modulated electroluminescence spectra and low frequency (≤ 100 kHz) impedance characteristics of electroluminescent diodes are studied. Voltage modulated light emission tracks the onset of observed negative capacitance at a forward bias level for each modulation frequency. Active participation of sub-bandgap defect states in minority carrier recombination dynamics is sought to explain the results. Negative capacitance is understood as a necessary dielectric response to compensate any irreversible transient changes in the minority carrier reservoir due to radiative recombinations mediated by slowly responding sub-bandgap defects. Experimentally measured variations of the in-phase component of modulated electroluminescence spectra with forward bias levels and modulation frequencies support the dynamic influence of these states in the radiative recombination process. Predominant negative sign of the in-phase component of voltage modulated electroluminescence signal further confirms the bi-molecular nature of light emission. We also discuss how these states can actually affect the net density of minority carriers *available* for radiative recombination. Results indicate that these sub-bandgap states can suppress external quantum efficiency of such devices under high frequency operation commonly used in optical communication.

# I. Introduction

Presence of electronic defects is often a limiting factor for the performance of any semiconductor device[1-3]. Defects present in the active region of the device introduce sub-bandgap states which can alter the band edge density of minority charge carriers on either side of the junction. In electroluminescent devices, defects can degrade long term radiative efficiency and stability. Phenomenon of negative capacitance (NC), as a signature of inductive reactance[4, 5], has been reported in many semiconductor devices/structures like Schottky barriers[6, 7], solar cells[8, 9], Silicon diodes[10], p-i-n junctions[11], $TiO_2$ films[12], light emitting diodes (LED)[13, 14] and laser diodes (LD)[15] under high forward biases. In case of LEDs, onset of NC has been associated with the threshold of light emission[16] and for laser diodes, it has been correlated with the onset of lasing[15]. Voltage[16] and current[17] modulated electroluminescence at low modulation frequency (f $\leq$ 100 kHz) has also been studied in this context. In general, increase in modulated electroluminescence with high injection levels and low frequencies, has been qualitatively correlated with the observed NC which also follows the same trend. However, presence of NC has also been observed in case of Silicon based devices and solar cells, which are intrinsically inefficient for light emission. To sum up, occurrences of NC have always been associated with forward biased junctions and a variety of explanations have been proposed to justify its presence. Nevertheless, basic understanding of physical mechanisms of NC, which can be extendable to various kinds of devices, is required. In our study, we have tried to comprehend NC in a more generalized way by considering the role of sub-bandgap states in the light emission process. Reported variation of modulated electroluminescence with low frequency[16] also hints at this participation. How these sub-bandgap defects actually affect both NC and modulated electroluminescence in a significant way is still not very clear. Experimental investigations of the influence of sub-bandgap defects on the physics of NC as well as that of modulated electroluminescence in light emitting devices is the main focus of the present work. Moreover, we also extend and generalize our explanations of NC to non-electroluminescent devices. Better understanding of these interconnected phenomena can be used to characterize as well as to improve the performance of semiconductor devices.

In this study, we employed low frequency (f $\leq$ 1 MHz) impedance spectroscopy and voltage modulated electro-luminescence (VMEL) spectroscopy (f $\leq$ 100 kHz) at room temperature under small signal condition ($\Delta V_{ac}^{rms}$ = 30 mV) to investigate commercially available LDs and LEDs. Here we will report – (i) the onset of relatively dispersion less region in the measured steady state conductance soon after the growth of 'significant' continuous wave (CW) light output, (ii) a plausible physical mechanism for NC, (iii) connection between the onset of negative capacitance with the onset of VMEL signal at a particular modulation frequency and forward bias ($I_{dc}$) and (iv) contribution of sub-bandgap defects to the spectral line shapes of VMEL with the variation of injection level as well as the modulation frequency.

We will also discuss the importance of using the above characterization techniques like modulated light emission to optimize the material parameters and device designs. More specifically, we will argue that such optimizations are necessary to minimize the contribution of slowly responding defect states in light emission processes. This can significantly improve the performance of high speed (~GHz) operation of electroluminescent devices used in optical communication.

## II. Experimental Methods

For this study, we used commercially available diode laser DL 3148 from Sanyo (lasing wavelength 635 nm) at current injection levels below the lasing threshold. Results are also reproduced for a number of such DL-3148 LDs and other LDs like RLD-65-NE from Rohm (657 nm) as well as for red LEDs. A simple equivalent circuit model (inset, figure 1b) consisting of an apparent capacitance ($C_{app}$) in parallel with an apparent conductance ($G_{app}$) is used by neglecting series resistance at these current injection levels. For impedance measurements, we used E4980A precision LCR meter from Agilent ($f \leq 2$MHz) and the forward bias injection current level, as monitored by the same instrument, is referred as $I_{dc}$. In case of VMEL measurements, E4980A was used for dc-biasing and SR850 lock-in amplifier ($f \leq 100$ kHz) from Stanford Research System for voltage modulation. Electroluminescence signal was dispersed using an Acton Monocromator-2555i from Princeton Instruments ($\Delta\lambda = 1.85$ nm) and was finally detected with a Si photodiode (FDS010) in phase with the applied voltage modulation. The current signal from the photodiode was then amplified with a SR570 current preamplifier from Stanford Research System before being fed to SR850 lock-in amplifier for final measurement. Under small signal condition, measured $I_{ac}^{rms}$ was kept at least an order of magnitude smaller than $I_{dc}$ even at injection levels of a few mA. To couple the dc bias and ac modulation signal, we used an adder-amplifier circuit which limits $I_{dc}$ to less than 10 mA. For CW light detection, emitted light was externally chopped at 35 Hz using a SR540 chopper controller from Stanford Research Systems. We monitored the temperature of all samples using a Lakeshore 340 temperature controller along with DT-670B-SD Silicon diode sensor at all stages of bias currents.

## III. Results and Discussions

Figure 1a is the traditional L-I characterization of the laser diode DL-3148, showing variation of CW light emission with forward injection current level ($I_{dc}$). Here we see the usual lasing threshold around 32 mA. However, in all our further discussions we will only focus on the current levels below this lasing threshold. Inset of figure 1a shows onset of light emission for $I_{dc}$ ~2 μA. We also notice that light output grows significantly only after a forward current of ~20 μA. Figure 1b and 1c show variation of $G_{app}$ and $C_{app}$ respectively with $I_{dc}$ at different modulation frequencies (f). Two distinct frequency dependent regions for injection currents – (a) < 20 μA and (b) > 3 mA can be seen in $G_{app}$ which have qualitatively opposite frequency dependent nature. For $I_{dc}$ < 20 μA, the device is dominated by diffusion of injected minority charge carriers hence, $G_{app}$ increases monotonically with both the injection level and modulation frequencies. Whereas, we notice that $C_{app}$ increases with increasing injection level but decreases with increasing modulation frequency. This is characteristic diffusion capacitance like behavior where minority carrier storage near the junction plays a major role (chapter # 2, reference # 1). However, we notice a relatively non-dispersive region in $G_{app}$, triggered after significant light output, for 20 μA < $I_{dc}$ < 3 mA. Interestingly $C_{app}$ remains highly dispersive in this range, which starts to decrease and eventually turns negative with increasing $I_{dc}$. For lower frequencies, this NC is more prominent and observed at lower injection levels. Being real and imaginary parts of the same dielectric response function, qualitatively different frequency dependence of $G_{app}$ and $C_{app}$ may seem counterintuitive. In the context of Kramers-Kronig relations for linear dielectric responses[18], there should be a direct correlation between $C_{app}$ and $G_{app}/\omega$ ($\omega$ being angular frequency of modulation), which is also true in present measurements (not shown here). Hence the different behavior of $C_{app}$ and $G_{app}$ is not a cause of concern. A hump in both $G_{app}$ and $C_{app}$ has been observed near the lasing threshold for different modulation frequencies. Clearly frequency dependent NC can neither be caused by CW light emission which starts at quite low current (~2 μA) nor by the lasing action which starts at a comparatively higher current (> 30 mA). At this point, we postpone any further discussion on the cause of NC to the later part of this paper.

It is evident from the variation of $C_{app}$ with such low modulation frequencies (≤ 1 MHz) that sub-bandgap states with transition rates much slower than the band-to-band radiative transitions are playing a significant role. However, a dispersion free region for $G_{app}$ in the same range of currents points to the fact that the modulation at different frequencies couples the energy used in the injection process with the surrounding heat bath[18] in a similar fashion. Usually, Schockley-Read-Hall type of recombination, mediated by sub-bandgap trap states, dissipates heat[19] through lattice vibrations (joule heating ~ $i^2R$). So it is clear that sub-bandgap states can notably affect the reactive response under different modulation rates. Nevertheless their steady state dissipation of energy is insignificant in the intermediate current injection levels (figure 1b). Comparing figures 1a and 1b, it is also evident that this non-dispersive region in $G_{app}$ starts only after a 'significant' growth of CW light emission. In order to understand this event, we presume that the energy spent in increasing the forward bias and subsequent increase in minority carrier injection is wholly being utilized by the radiative light emission of those injected minority carriers without significant coupling the excess energy to the heat bath in the frequency range of operation. At higher current injection levels, increased joule heating is possibly significant enough to change this dispersion less situation and $G_{app}$ starts showing variations with modulation frequency for $I_{dc}$ > 3 mA. This is supported by the fact that the sample temperature remains flat around 298 ± 0.1 K for $I_{dc} \leq$ few mA and sharply increases thereafter. It is interesting to note the qualitative difference in frequency variation of $G_{app}$ below (diffusion capacitance like behavior) and above (significant joule heating) the relatively dispersion free region. However, we must also state that the overall span of this dispersion less regime with respect to the forward current levels in fact varies for different laser diodes.

To understand the physical mechanism of NC, we have also measured VMEL. Figure 1d shows variation of VMEL for DL 3148 with $I_{dc}$ (< 10 mA, as limited by the adder amplifier circuit) at different frequencies, which indicates: (i) signal increases with increasing bias, (ii) for lower frequency, onset of signal is at lower values of $I_{dc}$ and (iii) signal is larger for lower value of modulation frequency. Comparison between figures 1c and 1d tells us that *VMEL signal also starts to increase significantly around the current bias values where $C_{app}$ turns fully negative for a specific modulation frequency*. Low frequency response of VMEL again suggests participation of slowly responding defect states in the modulated light emission processes. Strong correlation of the onset of NC with that of the modulated light emission actually rules out any contact problems as one of the origin of negative capacitance. However, the magnitude of NC does not scale with the modulation frequency like a normal inductor even for f ~ 1 MHz, which also rules out[6] any contribution from external parasitic inductances.

Generally speaking, any inductive type of response originates when a sample builds resistance to changing current levels and modulated current lags behind the applied voltage modulation. Quasi Fermi levels which track the injected minority carrier densities (figure 2), are influenced by both radiative recombinations and contributions from sub-bandgap defect states.

Onset of radiative recombination irreversibly depletes minority carrier reservoirs stored on the either side of the junction with increasing forward bias. This reduction happens at a rate faster than the rate of replenishment of minority carriers by injection and by carrier trapping-de-trapping processes. This way, active participation of slowly responding sub-bandgap defects can actually delay the dynamic response of quasi Fermi levels. If this contribution of sub-bandgap states to overall density of *available* minority carriers is significant, then quasi Fermi levels cannot follow the applied modulation. Thus, steady state separation between these levels goes through a transient change over a full cycle of sinusoidal variation. Such transient changes further induce a transient variation of the steady state current level, which is proportional to the spatial gradient of these quasi Fermi levels. *As a consequence, small signal reactive response of the device acquires an inductive component which resists any such deviations from the steady state current level.* Resultant induced voltage ultimately tries to compensate any such momentary changes in the separation of quasi Fermi levels over a cycle of applied modulation and steady state condition is regained. The ultimate effect of this active compensation is that the transient current slowly dies and net current level approaches[4, 5] its steady state value as governed by $I_{dc}$. An increase in $I_{dc}$ increases the number of these injected minority carriers as well as enhances the amount of radiative recombination. This in turn requires larger 'inductive' like response to restore and maintain the steady state separation of quasi Fermi levels over a full cycle of voltage modulation. As a consequence, capacitance part of the reactance becomes more and more negative with increasing $I_{dc}$ as is evident in 1c. At this stage, we would like to add that any past observations[7, 10, 11] of such generic frequency dependence of NC in predominantly non-radiative Silicon based structures/devices may also require the presence of mutually competing 'fast' and ‘slow’ non-radiative channels near the active junction. Such dynamic competition between these ‘fast’ and ‘slow’ processes can generate transients as mentioned above. This can finally trigger a similar compensatory type negative capacitance response where modulated current will lag behind the modulating voltage like that of an inductor to negate any transient change of steady state current level. However, this still has to be quantitatively verified by experiments on a case by case basis.

To explain the variation of VMEL with current bias and modulation frequency as observed in figure 1d, we will now try to understand the mechanism by which sub-bandgap defect states contribute towards the net minority carrier density *available* for radiative recombination at a particular temperature (T) and modulation frequency (f). In connection with NC, we have already discussed the importance of slowly responding states which exchange carriers with the injected minority carrier reservoir at rates comparable or less than the frequency of applied modulation. These can be sub-bandgap states away from the band edges near the active junction. To further clarify, we present figure 2 which depicts a schematic energy (E) level diagram illustrating the contribution of defect states to the total minority carrier (here electrons) density $n_{injected}^{Total}(E)$ available at the conduction band edge ($E_C$) in the active region of the device. Here we assume that the electron quasi Fermi level ($E_{Fn}$) can be extended in the active region to describe the population of injected minority electrons. Similar diagrams can be made for minority holes too. As per the diagram, the total density of injected minority electrons $n_{injected}^{Total}(E)$ in the active region is the sum of free minority electron density $n_{injected}^{Free}(E)$ and the density of trapped minority electrons as

$$n_{injected}^{Total}(E) = n_{injected}^{Free}(E) + n_{injected}^{Trapped}(E) \quad (1)$$

Where, $n_{injected}^{Trapped}(E) = \int_{E_C - E_{Th}}^{E_{Fn}} g(E)dE$ and g(E) is the sub-bandgap defect density at an energy E below the conduction band. At a particular T (in Kelvin) and f (in Hz), $n_{injected}^{Trapped}(E)$ have contributions only from sub-bandgap defects which lie within $E_{Fn}$ and $E_{Th}$ (shaded area in figure 2). $E_{Th}$ is the characteristics thermal activation energy of sub-bandgap states, given by:

$$E_{Th} = k_B T \ln(\nu / f) \quad (2)$$

If the rate at which these states can exchange carriers with the band edge is $\sim \frac{1}{\tau}$ then equation (2) can be rewritten as $\frac{1}{\tau} = \nu \exp\left(-\frac{E_{Th}}{k_B T}\right)$, where τ is the characteristic time, ν is the thermal prefactor of such transitions and $k_B$ is the Boltzman constant. With increasing forward bias, $E_{Fn}$ moves closer to the conduction band and the difference $(E_C - E_{Fn})$ decreases. Depending on T and f, defects will contribute to the steady state value of band edge carrier density only if $(E_C - E_{Th}) > (E_C - E_{Fn})$ . It is clear from equation 2 that sub-bandgap states responding at low frequencies stay away from the band edges and these deeper states can satisfy the above criterion at a smaller forward bias. As a result, in figure 1d, we see that the onset of VMEL signal as well as that of the negative capacitance response happens at smaller forward biases for lower frequencies. The shaded area in figure 2 represents those defect states which contribute toward $n_{injected}^{Trapped}(E)$. This $n_{injected}^{Trapped}(E)$ increases with increasing $E_{Th}$ and with decreasing f. Consequently, the net minority carrier density [ $n_{injected}^{Total}(E)$ ] *available* for radiative recombination process also increases with decreasing modulation frequency. Therefore, we notice in figure 1d that room temperature VMEL signal at a fixed forward bias is higher for lower modulation frequencies. Smaller the modulation frequency, larger is the contribution from deeper defects (having larger $E_{Th}$) which are also considerably slower in terms of their response time. That way the re-equilibration of minority carriers takes longer time and quasi Fermi level separation as well as current response lags the voltage modulation by a bigger margin. This in turn requires stronger inductive type of response to maintain the steady state. As a result, observed negative capacitance is much more negative for lower frequencies (figure 1c) at a particular $I_{dc}$. Strong dependence of radiative recombination on steady state contribution of minority carriers from sub-bandgap states may also signify that radiative and non-radiative recombinations of these minority carriers are no longer mutually exclusive events. Subsequently this may also imply $\frac{1}{\tau_{Effective}} \neq \frac{1}{\tau_{radiative}} + \frac{1}{\tau_{Non-radiative}}$. Here the equality sign applies only in case of non-interconnected probabilistic processes assuming a particular minority carrier cannot take part in both radiative and non-radiative processes which is not strictly true under the current circumstances. Moreover, any major contribution of $n_{injected}^{Trapped}(E)$ towards the total density of minority carriers (equation 1) *available* for radiative recombination can have significant impact on the efficiency of any electroluminescent device. Minority carriers trapped by existing deep defects will not respond at high frequency (~GHz) and will be missing in action from the final radiative recombination processes as per equation 1 and figure 2. This can effectively suppress the amount of minority carriers accessible for radiative recombination, thereby reducing the overall light intensity.

Therefore, these slower processes can, in principle, compromise high frequency applications of these light emitting device used in optical communications.

In figure 3a, we plot the CW light emission spectra of LD at different $I_{dc}$. Expected increase in the light emission intensity and progressive sharpening of the spectra with increasing injection current levels can easily be noticed. Inset of figure 3a shows the spectrum at an injection current above the lasing threshold and normal lasing action can be identified by its narrow spectral shape. In-phase VMEL spectra at different current injection levels are also plotted in figure 3b for a fixed f = 10 kHz. We notice that VMEL signal level increases at higher forward biases as expected. Shape of the measured VMEL spectra resembles the first derivative line shapes of modulated reflectance spectroscopy where the external modulation does not significantly perturb the translational invariance[20] of the solid. It seems that the voltage modulation also does not change the electric field of the forward biased junction enough to cause any field induced acceleration of the charge carriers in the active region. From this first derivative like spectral line shape, it is rather easy to conclude that most of the voltage modulation related effects come from changes in the minority carrier density. We can write the net rate of recombination when the system is driven away from the steady state as:

$$R_{Net} = \frac{(n+\Delta n)(p+\Delta p)}{np} R_{Steady} \approx R_{Steady} + \Delta R \qquad (3)$$

where n, p are steady state densities of injected minority electrons and minority holes respectively, $R_{Steady}$ is the generic van Roosbroeck-Schokley type recombination rate for electroluminescence in the steady state and ΔR is the deviation of radiative recombination rate from the steady state. After neglecting small second order terms, it can be shown (following chapter 6 of reference # 3) that:

$$\frac{\Delta R}{R_{Steady}} \approx \frac{\Delta n}{n} + \frac{\Delta p}{p} \quad (4)$$

Assuming $\Delta n = \Delta p = \Delta c$, one can arrive at:

$$\Delta R \sim \Delta c \quad (5)$$

Where Δc is the excess minority carrier density along with a proportionality constant consisting of only steady state parameters. Therefore, any increase/decrease in minority carrier density (Δc) following the voltage modulation (ΔV) can proportionately modulate the rate of radiative recombination as well as the resultant electroluminescence intensity. It is also evident from figure 2 that modulating the separation of quasi Fermi levels with the modulating voltage actually affects the amount of minority carrier *available* for band-to-band recombination. Here we are assuming that there is no significant pinning of quasi Fermi levels for both types of minority carriers. It is equivalent to modulating the shaded area in figure 2, between $E_{Fn}$ and $E_{Th}$ by modulating the quasi Fermi level position. Hence, these types of modulated changes in VMEL are due only to modulation of the pumping levels as illustrated in figure 4a.

Moreover, it is evident from figure 1c that reactive response at a particular modulation frequency is affected strongly by the change in the net injection level. Therefore, it won't be wrong to assume that voltage modulation induced perturbation of minority carrier density not only modulates the radiative recombination but also concurrently varies the dielectric response of the diode. We also note that the measured in-phase VMEL signal (figure 3b) becomes increasingly negative as the spectral scan goes through its peak. We have earlier pointed out the connection between the onset of VMEL signal with the onset of NC at a particular modulation frequency (figure 1c and 1d), whereas, any such connection between the measured conductance data (figure 1b) and onset of VMEL signal is missing. Since the steady state reactive response of minority electrons (n) and minority holes (p) depends on respective densities, therefore, VMEL signal is directly proportional to the steady state reactive response of both, minority electrons and minority holes, as VMEL ~ R ~ np. We know that the time varying current through a reactive component is always out-of-phase with the small signal voltage modulation. Therefore, prevailing reactive contributions at high forward biases for both minority electrons $\left(n \sim e^{\frac{i\pi}{2}}\right)$ and minority holes $\left(p \sim e^{\frac{i\pi}{2}}\right)$ combined together can produce the overall negative factor $\left[VMEL \sim R \sim np \sim e^{\frac{i\pi}{2}} \times \sim e^{\frac{i\pi}{2}} \cong e^{i\pi} = -1\right]$ of the in-phase VMEL signal. This characteristic negative sign of the in-phase VMEL is, therefore, a testimony to the bi-molecular nature of the light emission process.

In figure 4b, we plot the VMEL (in-phase) spectra for different modulation frequencies at a fixed forward current bias ($I_{dc}$) of 1 mA. Here we see the increase in VMEL signal level with decreasing modulation frequency as also explained in connection with figure 1d. Strong dependence of VMEL lineshape on modulation frequency is a clear sign of the active presence of defect states in modulating the total injected minority carrier density at the band edges inside the active region and consequently affecting their radiative recombination. It is likely that injected minority carriers at a particular forward bias easily get trapped and subsequently re-emitted by the sub-bandgap defects at time scales longer than few tens of microseconds. As evident from the band diagram in figure 2, lowering the modulation frequency scans a wider range of energy depth ($E_{Th}$). Therefore, we actually see a cumulative contribution of defect states located at different energy depths below the quasi Fermi level. As a result, in-phase VMEL signal also increases with decreasing modulation frequency. It has been reported[21, 22] earlier that it is necessary to consider the modulation frequency dependence of only the in-phase component of the photoreflectance signal (third derivative line shape) to appreciate the detailed role of sub-bandgap defects in the context of modulation spectroscopy. Further work is going on to comprehend the physics of modulation spectroscopy of electroluminescence. However, it is not necessary that the theoretical analyses[21, 22] widely used for modulated reflectance spectra of a reverse biased junction depleted of free carriers will also be valid for modulated electroluminescence spectra of a device experiencing strong radiative recombinations of injected minority carriers under large forward biased currents.

## IV. Conclusion

In summary, we have shown that negative capacitance is not associated directly with the CW lasing threshold or with the onset of CW light emission in an electroluminescent device. Rather, we perceive it as a result of significant participation of sub-bandgap states in the active region towards the overall radiative recombination process. Dependence of negative capacitance and voltage modulated electroluminescence on low modulation frequencies also confirms the participation of sub-bandgap states in the minority carrier dynamics even at low injection levels. We can relate the onset of negative capacitance with the presence of ‘inductive’ like response necessary to compensate any transient effect arising out of a dynamic competition between the ‘fast’ radiative recombination pathways and the slowly responding electronic defects. Device designs should minimize any considerable contribution from such slower carrier trapping-de-trapping processes toward the overall density of minority carriers *available* for radiative recombination. Only then can we expect a noteworthy enhancement in the intensity of emitted light modulated at very high frequencies (~GHz) necessary for applications like optical communication. A qualitatively similar consideration may also apply for high frequency applications of Silicon based devices. Behavior of in-phase VMEL signal connects the dielectric properties of the junction with the bimolecular nature of light emission in electroluminescent devices. Studies on temperature dependence of VMEL and time evolution of current transients are going on to develop more quantitative details of the role of sub-bandgap states towards modulated electroluminescence spectra. These studies are aimed at improving the efficiency and long term stability of these light emitting devices and definitely need further attention.

## Acknowledgements

Authors want to thank IISER-Pune for funding the laboratory infrastructure. Authors also wish to thank Prof. B. M. Arora for valuable discussions and Prof. K. N. Ganesh for support. KB is thankful to CSIR, India for research fellowship.

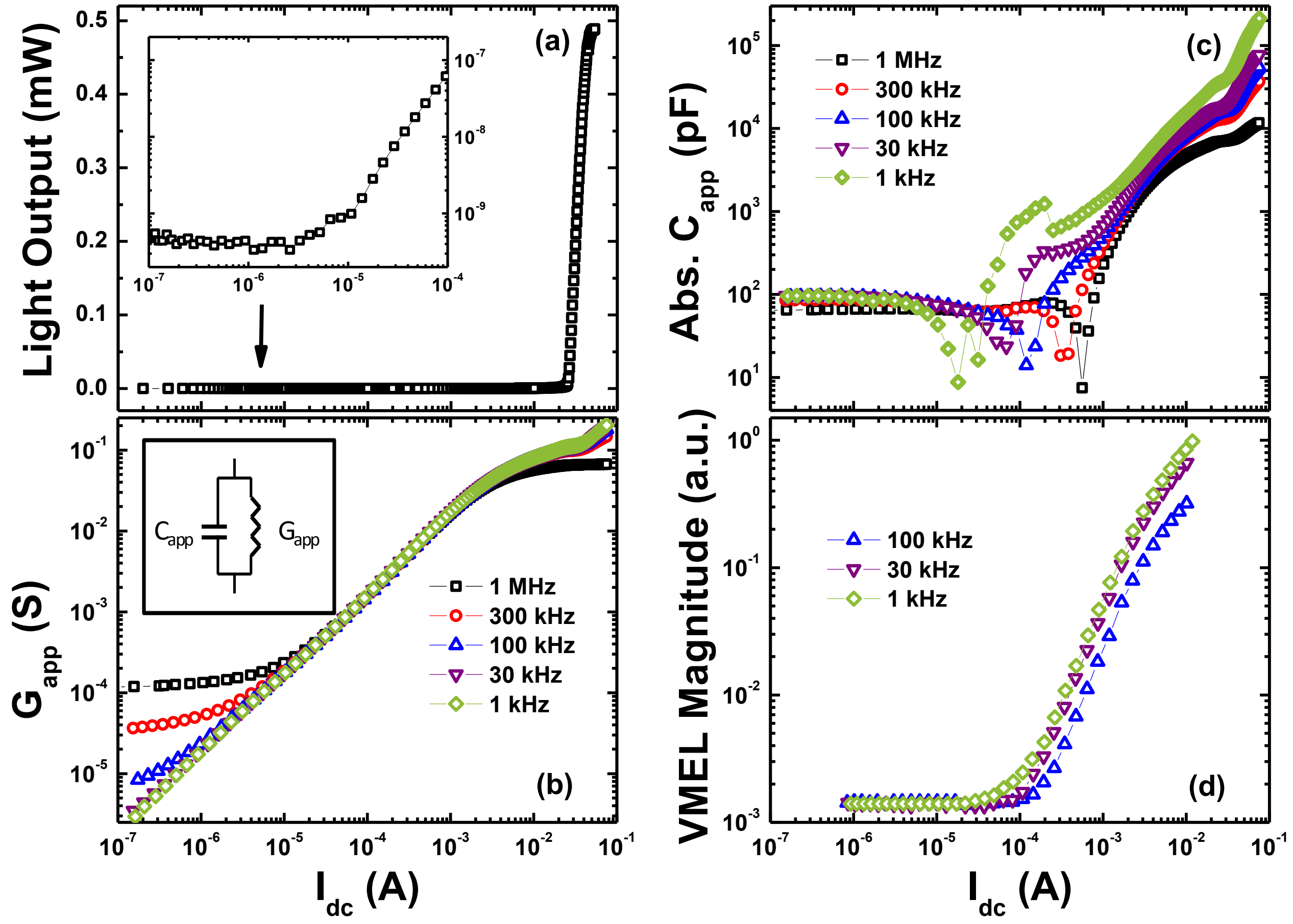

**Figure 1:** (a) CW light output vs forward current bias $I_{dc}$, showing laser threshold around 32mA. Inset shows the onset of CW light output at a lower $I_{dc}$. (b) Conductance of the light emitting device vs $I_{dc}$ at different frequencies, showing two distinct dispersive regions separated by a relatively non-dispersive region. Frequency dependences of dispersive regions are characteristically opposite. Span of non-dispersive region varies for different samples. Inset shows the simple equivalent circuit model used in our analysis. (c) Capacitance vs $I_{dc}$, plotted as absolute value, turning of the curve after sudden kink is the starting of negative capacitance (d) VMEL signal vs $I_{dc}$. Onset of VMEL signal at a particular frequency matches with the onset of negative capacitance.

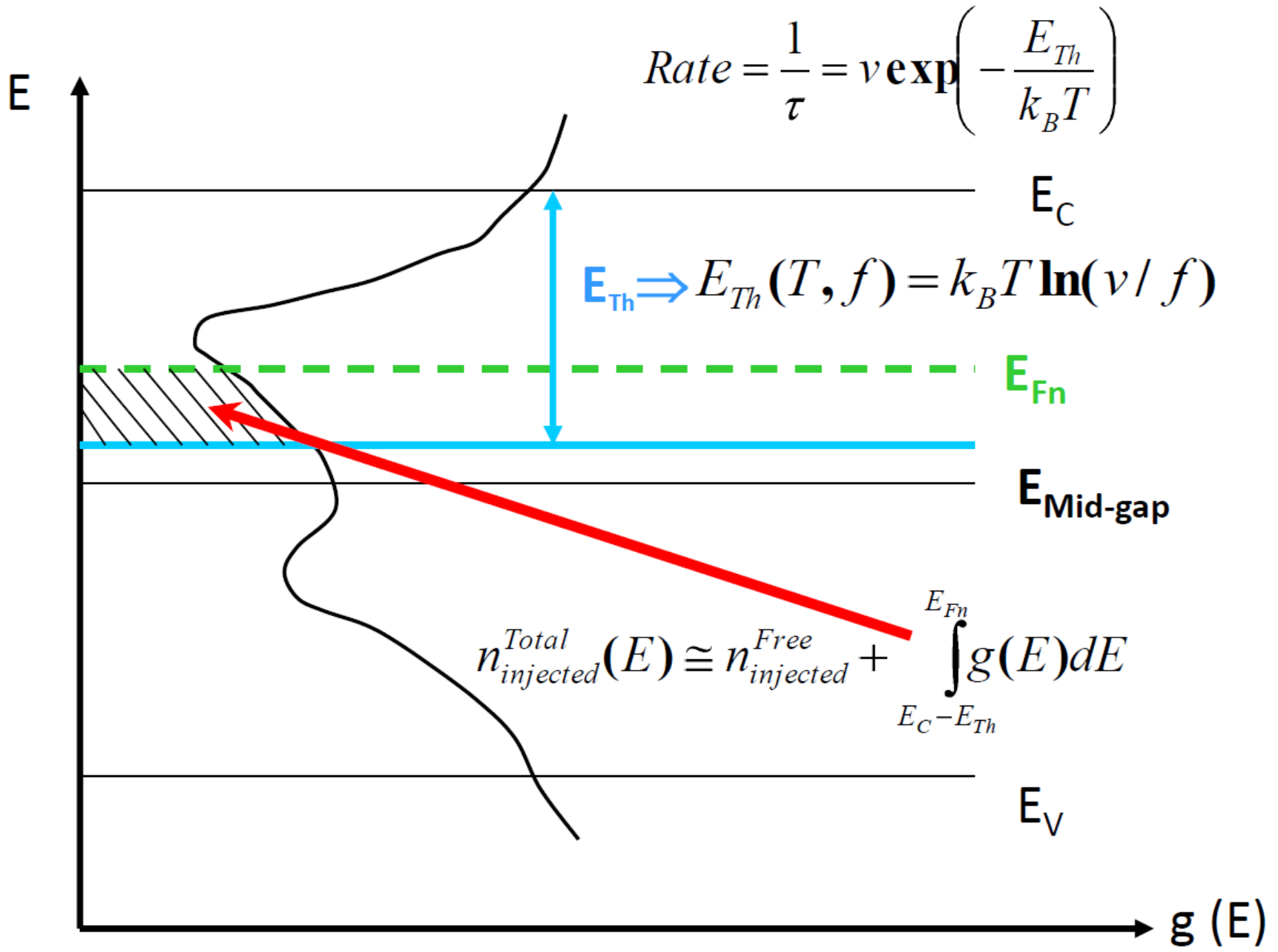

**Figure 2:** Effective band diagram to understand the contribution of sub-bandgap defects to the injected minority carriers (electrons in the p-type side of the junction) *available* for recombination at a particular temperature and frequency. Here we have assumed that the quasi Fermi level ($E_{Fn}$) can be extended to describe the population of injected minority electrons near the active region. The shaded area represents the energy range of the sub-bandgap states which are actively contributing minority carriers for band-to-band recombination. Based on a zero Kelvin picture, states below $E_{Th}$ cannot follow the modulation and cannot contribute to minority carrier density at the band edge.

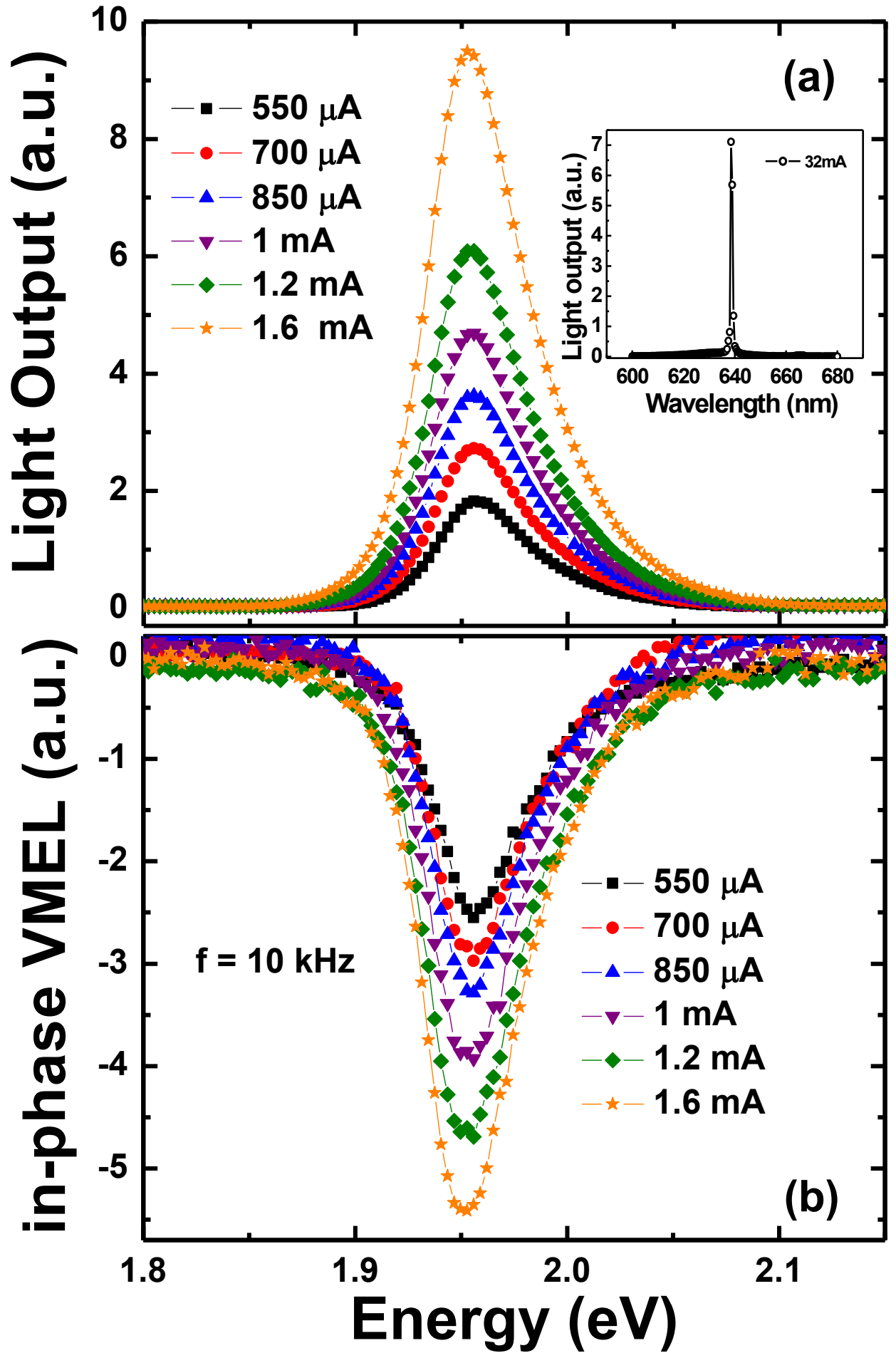

**Figure 3:** (a) The usual CW electroluminescence spectra from DL 3148 red laser diode below the lasing threshold. Inset shows the sharp spectral line shape just above the lasing threshold. (b) In-phase VMEL spectra at different forward bias currents below the lasing threshold. The in-phase VMEL signal progressively gains a distinct negative component near its peak for all bias levels shown in the figure.

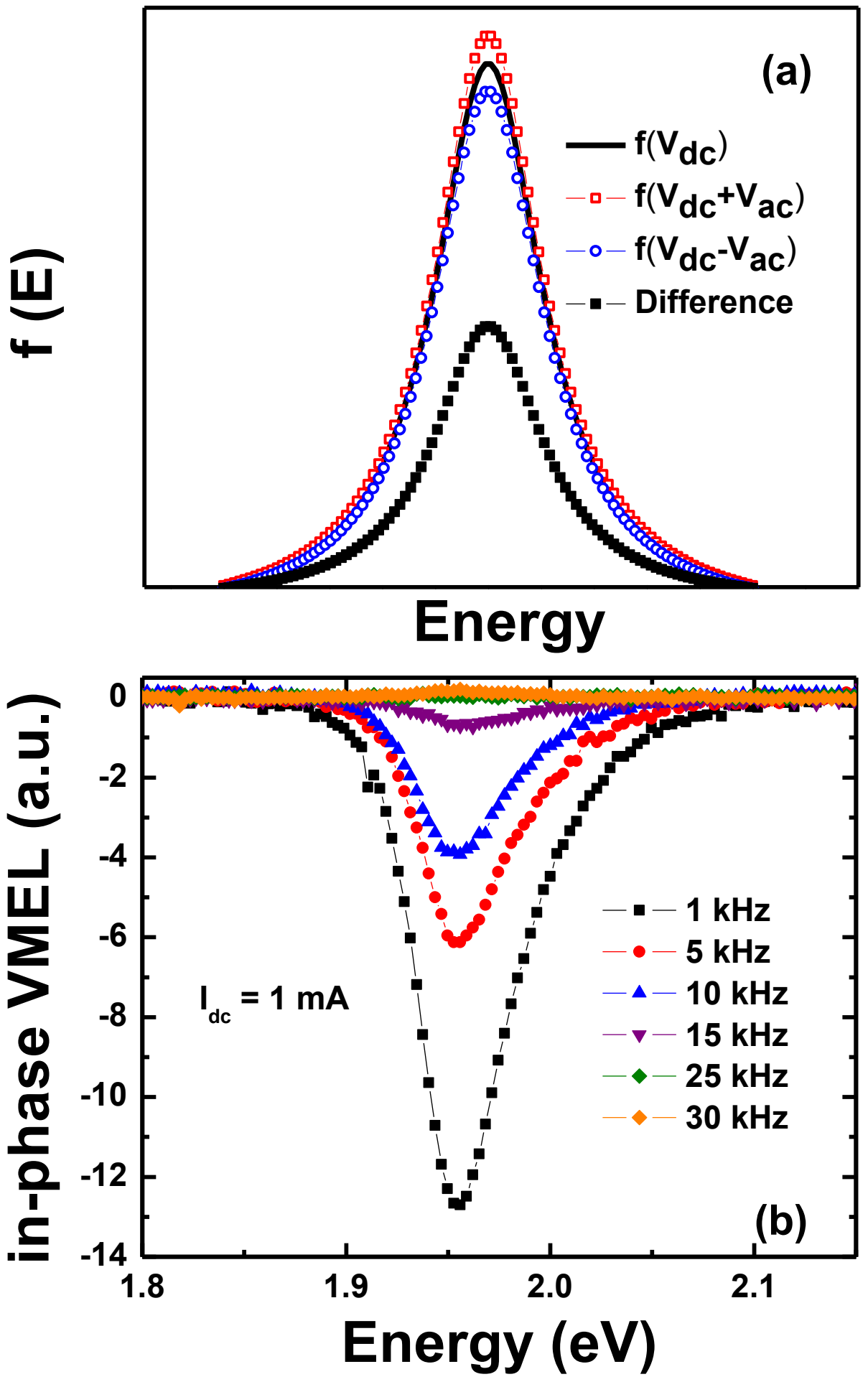

**Figure 4:** (a) Schematic of CW VMEL spectra (solid line) as modulation of pumping levels (Δn and Δp) by small signal voltage modulation (empty symbols). Absolute difference between modulated and CW spectra (solid symbol) (b) Variation of in-phase VMEL spectra with modulation frequency. Lower frequencies produce cumulatively larger VMEL signal as explained by figure2.